\documentclass{article}
\usepackage{amsmath, amsthm, amssymb}
\usepackage{epsfig}
\RequirePackage{amsfonts}
\usepackage{longtable}
\usepackage{bbold}

\usepackage{fullpage}
\newtheorem{theorem}{Theorem}[section]

\theoremstyle{definition}
\newtheorem{definition}[theorem]{Definition}

\allowdisplaybreaks[1]

\newcommand{\fullsub}[4]{
\begin{array}{ccc}
\vspace{-6pt}{#1}\hspace{-9pt}&{#2}&\hspace{-9pt}{#4}\\
&{}_{#3}&
\end{array}
}

\begin{document}

\begin{center}{\Large \textbf{A method to calculate correlation functions for $\beta=1$ random matrices of odd size}}\\ \vspace{36pt}{\large Peter J. Forrester and Anthony Mays}\\ \vspace{18pt}\textit{Department of Mathematics and Statistics, University of Melbourne, Victoria 3010, Australia}
\end{center}

\vspace{36pt}

\begin{abstract}
The calculation of correlation functions for $\beta=1$ random matrix ensembles, which can be carried out using Pfaffians, has the peculiar feature of requiring a separate calculation depending on the parity of the matrix size $N$. This same complication is present in the calculation of the correlations for the Ginibre Orthogonal Ensemble of real Gaussian matrices. In fact the methods used to compute the $\beta=1$, $N$ odd, correlations break down in the case of $N$ odd real Ginibre matrices, necessitating a new approach to both problems. The new approach taken in this work is to deduce the $\beta=1$, $N$ odd correlations as limiting cases of their $N$ even counterparts, when one of the particles is removed towards infinity. This method is shown to yield the correlations for $N$ odd real Gaussian matrices.
\end{abstract}
\newpage
\section{Introduction}

\subsection{Background}

The identification of the statistics of a complicated physical system with the statistics of random matrices was first made explicit by Eugene Wigner \cite{wigner1955,wigner1957}. Motivated particularly by the difficulty of applying an individual particle model to the calculation of nuclear energy levels \cite{lane_thomas_wigner1955}, he studied the eigenvalues and eigenvectors of an ensemble of real symmetric matrices whose entries were normally distributed.

Due to concerns with defining a unique Gaussian ensemble and the difficulty of carrying out the mathematics, Dyson \cite{dyson1970} studied Circular Ensembles, where the matrix elements are points on the unit circle. In Dyson's formulation, he hypothesised identifying the behaviour of a sequence of $n$ eigenvalues (from a total of $N\gg n$ eigenvalues, all lying on the unit circle), with the behaviour of $n$ energy levels in a physical system. As part of the analysis, he developed the seminal distinction between Orthogonal, Unitary and Symplectic ensembles by examining time-reversal and rotational symmetries \cite{dyson1962a} --- the required symmetries imply that the respective ensemble be invariant under orthogonal (COE), unitary (CUE) and symplectic (CSE) transformations.

The analogous ensembles in the Gaussian case are the GOE, GUE and GSE. The required symmetries mean that GOE consists of real symmetric matrices, GUE of Hermitian complex matrices, and GSE of self-dual real quaternion matrices. As a consequence of these facts the eigenvalues of all the Gaussian ensembles are constrained to lie on the real line; the joint probability density functions (jpdfs) are \cite{mehta_and_dyson1963}
\begin{equation}
\label{eqn:GEjpdf}
P_{N,\beta}(x_1,...,x_N)=\frac{1}{C_{N,\beta}}\hspace{3pt}e^{-\beta\sum_{k=1}^N x_k^2/2}\hspace{3pt}\prod_{i<j}|x_i-x_j|^{\beta}
\end{equation}
with $\beta=1,2,4$ corresponding respectively to the GOE, GUE and GSE; $C_{N,\beta}$ is some normalisation constant.

Compare this with the jpdf for the circular ensembles:
\begin{equation}
\label{eqn:CEjpdf}
Q_{N,\beta}(\theta_1,...,\theta_N)=\frac{1}{C_{N,\beta}}\prod_{i<j}|e^{i\theta_i}-e^{i\theta_j}|^{\beta}
\end{equation}
where $\beta=1,2,4$ correspond respectively to COE, CSE and CSE. $C_{N,\beta}$ is again used to denote the normalisation. The eigenvalues of these ensembles lie on the unit circle in the complex plane. Note that (\ref{eqn:GEjpdf}) and (\ref{eqn:CEjpdf}) share the common feature that the interaction between the eigenvalues is precisely the distance between them raised to the power of $\beta$.

The clear distinction between the elements of the matrices of each Gaussian ensemble led Ginibre \cite{ginibre1965} to a generalisation specified by removing the symmetric\slash Hermitian\slash self-dual requirement. Despite the irrelevance of the orthogonal, unitary and symplectic terms in relation to their formation, these ensembles are named for their eponymous Gaussian cousins: GinOE of real matrices; GinUE of complex matrices; and GinSE of real quaternion matrices.

Ginibre's ensembles provided the motivation for the current work. GinUE is relatively straightforward; Ginibre himself worked out the the jpdf and eigenvalue correlations in his original paper of 1965. GinSE was slightly more opaque; Ginibre was able to identify the jpdf but was not able to find the correlations. Further work has since developed these correlations \cite{kanzieper2001}. On the other hand, GinOE proved quite intractable until very recently. Ginibre was only able to find the jpdf in the case where all eigenvalues are real. The full jpdf was not developed until 1991 \cite{lehmann_and_sommers1991}, and the complete correlations took longer still \cite{sommers_and_w2008}.

The trouble largely stems from there being different sectors to the jpdf, determined by the number of real eigenvalues \cite{lehmann_and_sommers1991,edelman1997,shukla2001}. Yet there is a further problem.

For an $N\times N$ matrix, the $N$ even and $N$ odd cases require a different analysis. In the $N$ even case the eigenvalue correlations have been the subject of the recent works \cite{b&s2007,forrester and nagao2007,sommers2007,sommers_and_w2008}. Of these, only \cite{sommers_and_w2008} gives an analysis of the correlations for $N$ odd. The feature that $N$ odd and $N$ even must be treated separately is already present in the calculation of the correlations for (\ref{eqn:GEjpdf}) and (\ref{eqn:CEjpdf}) in the case $\beta=1$. It is the aim of this paper to show that the calculation of these latter correlations, and those of GinOE, can be accomplished as a limit of the $N$ even cases.

\subsection{Parity problems for $\beta=1$}

For $\beta=1$ in (\ref{eqn:GEjpdf}) or (\ref{eqn:CEjpdf}) the general $n$-point correlation is given by an $n\times n$ quaternion determinant, or equivalently a $2n\times 2n$ Pfaffian (these are revised in Section \ref{Sec:oddTrouble}). This fact was revealed by Dyson \cite{dyson1970} in the case of (\ref{eqn:CEjpdf}) and further developed in a series of papers by Mehta \cite{mehta1971,mehta1976,mahoux_and_mehta1991}. The strategy of Dyson was to express the jpdf itself as an $N\times N$ quaternion determinant, and then to successively integrate this to deduce the $n$-point correlation. Successive integrations require a certain skew-orthogonality property that effectively reduces the size of the quaternion upon each integration. It is the first step which requires that $N$ even and $N$ odd be treated separately. In the method of Dyson the identity between the jpdf and the quaternion determinant involves a matrix factorisation; in the $N$ odd case this was achieved by the introduction of an auxiliary parameter $\delta$, and the limit $\delta\rightarrow 0$ is taken.

These complications make $N$ odd technically more difficult. In fact, the study of the correlations for a generalisation of (\ref{eqn:GEjpdf}) with $\beta=1$
\begin{equation}
\label{eqn:generalOEjpdf}
\frac{1}{C_N}\prod_{l=1}^Ne^{-V(x_l)}\prod_{1\leq j < k \leq N}|x_k-x_j|
\end{equation}
for general $V(x)$, given in \cite{mahoux_and_mehta1991}, was restricted to $N$ even. The modification required for $N$ odd is implied in the later paper of Frahm and Pichard \cite{frahm_and_pichard1995}, and the resulting formulae were written out explicitly in \cite{afnvm2000}.

The method of integration over alternate variables \cite{mehta1960}, or use of integration formulae due to de Bruijn for the product of a Pfaffian times a determinant \cite{debruijn1955} (each of which implies distinct treatment of even and odd), can be used to express the generating function for the correlations as an $N\times N$ Pfaffian. In the $N$ even case de Bruijn's method was used by Tracy and Widom \cite{tracy_and_widom1998} to obtain a formula for the generating function as the square root of the Fredholm determinant of a $2\times 2$ matrix integral operator, and functional differentiation is used to extract the correlations. Using essentially this same method Borodin and Sinclair computed the correlations for GinOE in the case of $N$ even.

For the case of $N$ odd, a quite different method, using Grassmannians and without the use of skew-orthogonal polynomials, was developed in \cite{sommers_and_w2008}. By introducing an artificial Grassmannian in the odd case, and using a diagrammatic method of expansion, the $n$th order GinOE correlations are generated for general $N$.

\subsection{Guide to paper}

Section 2 introduces the terminology and concepts required for the study of $\beta=1$ cases, and outlines the problems encountered when dealing with an odd number of eigenvalues. The difficulties that have postponed the identification of the jpdf and the calculation of the correlation functions for the GinOE are described, along with some of the attempts to overcome them.

Section 3 details the process of developing $N$ odd correlations for the jpdf (\ref{eqn:generalOEjpdf}) as limiting cases of the extant $N$ even correlations \cite{afnvm2000}. See also \cite{forrester?}.

In Section 4, the method as presented in Section 3 is applied to the GinOE using the existing $N$ even solution. The final section discusses the conditions under which this method would be applicable and suggests a future application.

\section{The trouble with being odd}
\label{Sec:oddTrouble}

Dyson \cite{dyson1970} identified that the correlation functions for $\beta=1,4$ are naturally expressed as quaternion determinants of matrices with quaternion elements. The problem can also be looked at from a Pfaffian viewpoint, since they are equivalent. This is the essence of the parity problems; Pfaffians of odd dimension are trivially zero, and their machinery does not seem immediately adapted to dealing with this.

We first look at the relevant definitions.

\subsection{Quaternion determinants and Pfaffians}
\label{subsec:QDet and Pf}

For a quaternion expressed as a $2 \times 2$ matrix
\begin{equation}
M_{i,j}=\left[\begin{array}{cc}
a & b\\
c & d\\
\end{array}\right]
\end{equation}
the dual of $M_{i,j}$ is written as
\begin{equation}
\bar{M}_{i,j}=\left[\begin{array}{cc}
d & -b\\
-c & a\\
\end{array}\right]
\end{equation}
A matrix $M$ consisting of $N \times N$ blocks of quaternions is said to be self-dual if
\begin{equation}
M_{j,i}=\bar{M}_{i,j}
\end{equation}

For ease of reference we provide here the definitions of quaternion determinants and Pfaffians.
 
\begin{definition} \textbf{Quaternion determinants}

Let $M$ be an $N \times N$ self dual matrix of quaternions $M_{i,j}$. The quaternion determinant is defined by
\begin{equation}
\mathrm{QDet}[M]=\sum_{P\in S_N}(-1)^{N-l}\prod_1^l(M_{ab}M_{bc}...M_{sa})^{(0)}
\end{equation}
The superscript $(0)$ denotes the operation $\frac{1}{2}\mathrm{Tr}$ of the quantity in brackets. $P$ is any permutation of $(1,...,N)$ which consist of $l$ disjoint cycles of the form $(a\rightarrow b \rightarrow c \rightarrow \cdot\cdot\cdot \rightarrow s \rightarrow a)$. If the $M_{i,j}$ are scalars then $\mathrm{QDet}[M]=\mathrm{Det}[M]$.

\end{definition}

\begin{definition} \textbf{Pfaffians}
\label{def:pfaff}

Let $X=[\tau_{ij}]_{i,j=1,...,2N}$ where $\tau_{ji}=-\tau_{ij}$ so that $\mathrm{X}$ is an antisymmetric matrix of even degree. Then the Pfaffian of $\mathrm{X}$ is defined by
\begin{eqnarray}
\nonumber \mathrm{Pf} [X]&=&\sum^*_{P(2l)>P(2l-1)}\varepsilon (P) \tau_{P(1),P(2)}\tau_{P(3),P(4)}\cdot\cdot\cdot\tau_{P(2N-1),P(2N)}\\
&=&\frac{1}{2^NN!}\sum_{P\in S_{2N}}\varepsilon (P) \tau_{P(1),P(2)}\tau_{P(3),P(4)}\cdot\cdot\cdot\tau_{P(2N-1),P(2N)}
\end{eqnarray}
where $\varepsilon (P)$ is the sign of the permutation. The * above the first sum indicates that the sum is over distinct terms only (that is, all permutations of the pairs of indices are regarded as identical).
\end{definition}

Usefully, Pfaffians can be calculated using a form of Laplace expansion. For a determinant, recall that you may expand along any row or column. For example, expand a matrix $A=[a_{ij}]_{i,j=1,...n}$ along the first row:
\begin{equation}
\mathrm{Det}[A]=a_{1,1}\mathrm{Det}[A]^{1,1}-a_{1,2}\mathrm{Det}[A]^{1,2} + \cdot\cdot\cdot (-1)^{n+1} a_{1,n}\mathrm{Det}[A]^{1,n}
\end{equation}
where $\mathrm{Det}[A]^{i,j}$ means the determinant of the matrix left over after deleting the $i$th row and $j$th column.

The analogous expansion for a Pfaffian involves deleting \textit{two} rows and \textit{two} columns each time. For example, expanding a skew-symmetric matrix $B=[b_{ij}]_{i,j=1,...n}$ ($n$ even) along the first row:
\begin{equation}
\mathrm{Pf}[B]=b_{1,1}\mathrm{Pf}[B^{1,1}]-b_{1,2}\mathrm{Pf}[B^{1,2}] + \cdot\cdot\cdot (-1)^n b_{1,n}\mathrm{Pf}[B^{1,n}]
\end{equation}
where $\mathrm{Pf}[B^{i,j}]$ means the Pfaffian of the matrix left after deleting the $i$th  and $j$th rows and the $i$th and $j$th columns. Laplace expansion requires $n!$ calculations for a determinant, and $n!!=n\cdot (n-2)\cdot (n-4 )\cdot\cdot\cdot$ in the case of a Pfaffian.

\subsection{A Pfaffian by any other name}

Define the matrix
\begin{equation}
Z_{2N}=\mathbf{1}_N\otimes \left[\begin{array}{cc}
0 & -1\\
1 & 0\\
\end{array}\right]
\end{equation}(that is, $Z_{2N}$ is a $2N \times 2N$ matrix consisting of $N$ $2 \times 2$ anti-symmetric blocks along the diagonal). By operating with $Z_{2N}$ we can change an $N \times N$ self-dual quaternion matrix into a $2N \times 2N$ anti-symmetric matrix. The effect of $Z_{2N}$ is to interchange each pair of columns and multiply every second column by $(-1)$.

Using $Z_{2N}$ we find the following simple relationship between Pfaffians and quaternion determinants
\begin{equation} \label{eqn:qdet=pf}
\mathrm{QDet} [M]=\mathrm{Pf}[MZ^{-1}_{2N}]
\end{equation}
For a proof of (\ref{eqn:qdet=pf}) see \cite{forrester?}.

In light of (\ref{eqn:qdet=pf}) it is understood that we may use the terms quaternion determinant and Pfaffian interchangeably.

\subsection{$2N$ or not $2N$}

Why is $N$ odd for $\beta=1$ special? This can best be understood by considering the generating function
\begin{equation}
\hat{Z}_{N,\beta}[a]:=\int_{\infty}^{\infty}dx_1\cdot\cdot\cdot\int_{\infty}^{\infty}dx_N\prod_{l=1}^Na(x_l)e^{-\beta V(x_l)}\prod_{1\leq j < k \leq N}|x_k-x_j|^{\beta}
\end{equation}

In the case $\beta=4$ the generating function can be expressed in the Pfaffian form \cite{mahoux_and_mehta1991}

\begin{equation}
\label{eqn:SEgenfn}
\hat{Z}_{N,4}[a]=N!\hspace{3pt}2^N\hspace{3pt}\mathrm{Pf}[\alpha_{j,k}]_{j,k=1,...,2N}
\end{equation}
where

\begin{equation}
\alpha_{j,k}=\frac{1}{2}\int_{\infty}^{\infty}e^{-2V(x)}a(x)\Bigl(Q_{j-1}(x)\tilde{Q}_{k-1}(x)-Q_{k-1}(x)\tilde{Q}_{j-1}(x)\Bigr)dx
\end{equation}
with $Q_n(x)$ an arbitrary monic polynomial of degree $n$ and
\begin{equation}
\tilde{Q}_{n-1}(x):=\frac{d}{dx}\Bigl(e^{-2V(x)}Q_{n-1}(x)\Bigr)
\end{equation}

In the case $\beta=1$ and $N$ even, the method of integration over alternate variables gives \cite{mahoux_and_mehta1991}
\begin{equation}
\label{eqn:OEgenfn}
\hat{Z}_{N,1}[a]=N!\hspace{3pt}2^{N/2}\hspace{3pt}\mathrm{Pf}[\gamma_{j,k}]_{j,k=1,...,N}
\end{equation}
where
\begin{equation}
\label{eqn:gammadef}
\gamma_{j,k}:=\frac{1}{2}\int_{\infty}^{\infty}dx \hspace{3pt}e^{-V(x)}a(x)R_{j-1}(x)\int_{\infty}^{\infty}dy \hspace{3pt}e^{-V(y)}a(y)R_{k-1}(y)\hspace{3pt}\mathrm{sgn}(y-x)
\end{equation}
with $R_n(x)$ an arbitrary monic polynomial of degree $n$. The structure of (\ref{eqn:SEgenfn}) is subtly, but crucially, different to (\ref{eqn:OEgenfn}) as the matrix in the latter is of even size only when $N$ is even, while that of the former is always of even size, regardless of the parity of $N$.

According to Definition \ref{def:pfaff} a Pfaffian is defined only for an even sized matrix. Extension of this definition to the odd-sized case has been given by de Bruijn \cite{debruijn1955} --- his treatment amounts to bordering by an additional column of $1$s and an additional row of $-1$s with a zero in the bottom right corner. In fact the method of integration over alternate variables also gives that the $N$ odd Pfaffian formula \cite{mehta2004} is obtained from the one for $N$ even (\ref{eqn:OEgenfn}) by bordering
\begin{equation}
\hat{Z}_{N,1}[a]=N!\hspace{3pt}2^{(N+1)/2}\hspace{3pt}\mathrm{Pf}\left[\begin{array}{cc}
[\gamma_{i,j}]_{i,j=1,...,N} & [\nu_i]_{i=1,...,N}\\
\left[\nu_j\right]_{j=1,...,N} & 0\\
\end{array}\right]
\end{equation}
where $\gamma_{i,j}$ is as in (\ref{eqn:gammadef}), while
\begin{equation}
\label{eqn:nudef}
\nu_i:=\frac{1}{2}\int_{-\infty}^{\infty}e^{-V(x)}a(x)R_{i-1}(x)dx
\end{equation}

The significance of the generating function is that the correlation functions follow by functional differentiation
\begin{equation} \label{eqn:fnal diff}
\rho_{(n)}(u_1,...,u_n)=\frac{1}{Z_N[a]}\frac{\delta^n}{\delta a(u_1)\cdot\cdot\cdot \delta a(u_n)}Z_N[a]{\Big |}_{a=1}
\end{equation}

Using (\ref{eqn:OEgenfn}) and (\ref{eqn:fnal diff}), Tracy and Widom \cite{tracy_and_widom1998} showed how the known $n\times n$ quaternion determinant formula from \cite{mahoux_and_mehta1991} for $\rho_{(n)}$ in the case $\beta=1$, $N$ even could be reclaimed. Crucial to their method is the identity
\begin{equation}
\label{eqn:1+ab}
\mathrm{Det}(\mathbb{1}_p+A_{p\times q}B_{q\times p})=\mathrm{Det}(\mathbb{1}_q+B_{q\times p}A_{p\times q})
\end{equation}
where $A_{p\times q}$ is a $p \times q$ (and $B_{q\times p}$ is a  $q\times p$) matrix valued integral operator. However, $\beta=1$ for $N$ odd has not yielded to this approach. Yet, the correlations in this case are known --- how then were they derived?

\subsection{Guessing the answer --- then verifying it}
\label{sec:guess}

For general jpdfs $P_{N,\beta}$, symmetric in all variables, the $n$-point correlation $\rho_{(n)}$ can be expressed as an integral according to
\begin{equation}
\label{eqn:integcorrelns}
\rho_{(n)}(r_1,...,r_n)=N(N-1)\cdot\cdot\cdot(N-n+1)\int dr_{n+1}\cdot\cdot\cdot \int dr_N \hspace{3pt}P_{N,\beta}(r_1,...,r_N)
\end{equation}
Note in particular that
\begin{equation}
\label{eqn:rho_recurrence}
\rho_{(n)}(r_1,...,r_n)=\frac{1}{N-n}\int dr_{n+1}\hspace{3pt}\rho_{(n+1)}(r_1,...,r_{n+1})
\end{equation}
This recurrence together with the initial condition
\begin{equation}
\label{eqn:rho_initial}
\rho_{(N)}(r_1,...,r_N)=N!\hspace{3pt}P_{N,\beta}(r_1,...,r_N)
\end{equation}
completely determines $\{ \rho_{(n)}\}_{n=1,2,...,N}$.

Now the functional differentiation method, when used in conjunction with a Pfaffian form of the generating function, is well suited for the computation of the $1$- and $2$- point correlations. Such a calculation does not rely on (\ref{eqn:1+ab}), and so can be applied to both the $N$ even and $N$ odd cases for $\beta=1$. It reveals that $\rho_{(n)}$ can, for $n=1,2$ be expressed as a $2n\times 2n$ Pfaffian with entries independent of $n$.

From this, one can then conjecture a Pfaffian form for $\rho_{(n)}$ in general. The validity of the conjecture requires verifying both (\ref{eqn:rho_recurrence}) and (\ref{eqn:rho_initial}). It is the verification of (\ref{eqn:rho_initial}) that requires $N$ odd be treated differently \cite{dyson1970,frahm_and_pichard1995}. The verification of (\ref{eqn:rho_recurrence}) is done by using a recursion for integrals of quaternion determinants, known as the Dyson Integration Theorem \cite{dyson1970,mehta1976,a&k2007}, and the parity of $N$ plays no explicit role.

\begin{theorem} {\rm \textbf{Dyson Integration Theorem}}
\label{thm:integral_identities}

Let $f(x,y)$ be a function of real, complex or quaternion variables where
\begin{equation}
\bar{f}(x,y)=f(y,x)
\end{equation}
with $\bar{f}$ being the function $f$, the complex conjugate of $f$ or the dual of $f$ depending on whether $x$ and $y$ are real, complex or quaternion respectively.

Also let 
\begin{eqnarray}
\label{eqn:dit1}\int f(x,x) d\mu (x)&=&c\\
\label{eqn:dit2}\int f(x,y)f(y,z) d\mu(y)&=&f(x,z)+\lambda f(x,z) -f(x,z)\lambda
\end{eqnarray}
for some suitable measure $d\mu$, a constant scalar $c$ and a constant quaternion $\lambda$.

Then for a matrix $F_{n\times n}=[f(x_i,x_j)]_{n\times n}$ we have
\begin{equation}
\int \mathrm{QDet}[F_{n\times n}] d\mu(x_n) = (c-n+1)\hspace{3pt}\mathrm{QDet}[F_{(n-1) \times (n-1)}]
\end{equation}
 
\end{theorem}
Remember that if $f(x_i,x_j)$ is a scalar, then QDet reduces to a standard determinant. For a proof of Theorem \ref{thm:integral_identities} see Theorem 5.1.4 in \cite{mehta2004}.

\vspace{26pt}The Pfaffian formula (\ref{eqn:OEgenfn}) involves an arbitrary set of monic polynomials $\{ R_j(x)\}_{j=0,1...}$. For the implementation of Dyson's integration formula, in particular the validity of (\ref{eqn:dit1}) and (\ref{eqn:dit2}), these polynomials must be chosen to have a certain skew-orthogonality property.

Define a skew-symmetric inner product $\langle \cdot | \cdot \rangle$ by
\begin{equation}
\langle f | g \rangle := \frac{1}{2}\int_{\infty}^{\infty}dx \hspace{3pt}e^{-V(x)}f(x)\int_{\infty}^{\infty}dy \hspace{3pt}e^{-V(y)}g(y)\hspace{3pt}\mathrm{sgn}(y-x)
\end{equation}
and let $\{R_n(x)\}_{n=0,1,...}$ be a corresponding family of monic skew orthogonal polynomials so that
\begin{eqnarray}
\langle R_{2m}|R_{2n+1}\rangle &=& -\langle R_{2n+1} | R_{2m}\rangle = r_n\delta_{nm}\\
\nonumber\\
\langle R_{2m} | R_{2n} \rangle&=&\langle R_{2m+1} |R_{2n+1}\rangle = 0
\end{eqnarray}

With this specification of $\{ R_j(x)\}_{j=0,1...}$ the correlations for the jpdf (\ref{eqn:generalOEjpdf}) have been shown, by verifying (\ref{eqn:rho_recurrence}) and (\ref{eqn:rho_initial}), to be given by explicit $n\times n$ quaternion determinant formulae. First the case when $N$ is even.

\begin{theorem}
\label{thm:general_even}
Let $\{R_n(x)\}_{n=0,1,...}$ be as above, then for an eigenvalue distribution given by (\ref{eqn:generalOEjpdf}), the $n$th order correlation function, in the case of $N$ even, is given by

\begin{equation}
\rho_{(n)}(x_1,...,x_n)=\mathrm{QDet}[f(x_i,x_j)]_{i,j=1,...,n}
\end{equation}
where
\begin{equation}
\label{eqn:generalOEkernel}
f(x,y)=
\left[\begin{array}{cc}
S(x,y) & \tilde{I}(x,y)\\
D(y,x) & S(y,x)\\
\end{array}\right]
\end{equation}
and
\begin{eqnarray}
\label{eqn:Seven}
S(x,y)&=&\sum_{k=0}^{N/2-1}\frac{e^{-V(y)}}{r_k}\Bigl( \Phi_{2k}(x)R_{2k+1}(y)-\Phi_{2k+1}(x)R_{2k}(y)\Bigr)\\
\nonumber\\
\label{eqn:Deven}D(x,y)&=&\sum_{k=0}^{N/2-1}\frac{e^{-V(x)-V(y)}}{r_k}\Bigl( R_{2k}(x)R_{2k+1}(y)-R_{2k+1}(x)R_{2k}(y)\Bigr)\\
\nonumber\\
\label{eqn:Ieven}\tilde{I}(x,y)&=&\sum_{k=0}^{N/2-1}\frac{1}{r_k}\Bigl( \Phi_{2k+1}(x)\Phi_{2k}(y)-\Phi_{2k}(x)\Phi_{2k+1}(y)\Bigr) + h(x,y)
\end{eqnarray}
\begin{eqnarray}
\Phi_k(x)&=&\int_{\infty}^{\infty}h(y,x)R_k(y)e^{-V(y)}dy\\
\nonumber \\
h(x,y)&=&\frac{1}{2}\mathrm{sgn}(y-x)
\end{eqnarray}

\end{theorem}

\vspace{26pt}For $N$ odd, the correlations exhibit the same functional form but with some modifications to the kernel (\ref{eqn:generalOEkernel}) as follows.

\begin{theorem}
\label{thm:general_odd}
\begin{eqnarray}
\nonumber S^{odd}(x,y)&=&\sum_{k=0}^{(N-1)/2-1}\frac{e^{-V(y)}}{\hat{r}_k}\Bigl( \hat{\Phi}_{2k}(x)\hat{R}_{2k+1}(y)-\hat{\Phi}_{2k+1}(x)\hat{R}_{2k}(y)\Bigr)\\
 &&+ \frac{e^{-V(y)}}{\hat{r}_{(N-1)/2)}}F(x)R_{N-1}\\
\nonumber\\
D^{odd}(x,y)&\hspace{-6pt}=&\hspace{-18pt}\sum_{k=0}^{(N-1)/2-1}\frac{e^{-V(x)-V(y)}}{\hat{r}_k}\Bigl( \hat{R}_{2k}(x)\hat{R}_{2k+1}(y)-\hat{R}_{2k+1}(x)\hat{R}_{2k}(y)\Bigr)\\
\nonumber\\
\tilde{I}^{odd}(x,y)&\hspace{-6pt}=&\hspace{-18pt}\sum_{k=0}^{(N-1)/2-1}\frac{1}{\hat{r}_k}\Bigl( \hat{\Phi}_{2k+1}(x)\hat{\Phi}_{2k}(y)-\hat{\Phi}_{2k}(x)\hat{\Phi}_{2k+1}(y)\Bigr) + h(x,y)\\
\nonumber\\
&&+\frac{1}{\hat{r}_{(N-1)/2}}\Bigl( \hat{\Phi}_{N-1}(x)F(y)-F(x)\hat{\Phi}_{N-1}(y) \Bigr)
\end{eqnarray}
with
\begin{eqnarray}
\hat{r}_n&:=&r_n \hspace{12pt} (n=0,...,(N-3)/2)\\
\nonumber\\
\hat{r}_{(N-1)/2}&:=&\int_{\infty}^{\infty}e^{-V(y)}F(y)R_{N-1}(y)dy\\
\nonumber\\
\hat{R}_n(x)&:=&R_n(x)-\frac{R_{N-1}(x)}{r_{(N-1)/2}}\int_{\infty}^{\infty}e^{-V(y)}F(y)R_n(y)dy\\
\nonumber\\
\hat{R}_{N-1}(x)&:=&R_{N-1}(x)\\
\nonumber\\
\hat{\Phi}_n(x)&:=&\int_{\infty}^{\infty}e^{-V(y)}h(y,x)\hat{R}_n(y)dy\\
\nonumber\\
F(x)&=&\frac{1}{2}
\end{eqnarray}

\end{theorem}
Proofs of these theorems, using this notation, are found in \cite{forrester?}. The original proofs are contained in \cite{mahoux_and_mehta1991,frahm_and_pichard1995,afnvm2000}.

\vspace{26pt}A problem presents itself if the jpdf is not a symmetric function in all variables. Then the recursion (\ref{eqn:rho_recurrence}) does not hold and so the strategy leading to Theorems \ref{thm:general_even} and \ref{thm:general_odd} is not valid. As mentioned, there is an alternative strategy involving the functional differentiation formula (\ref{eqn:fnal diff}). It however has not been successfully implemented in the case of $\beta=1$, $N$ odd.

As to be revised in the next section, the GinOE, $N$ odd case falls prey to both of these negative results, and so requires investigation of \textit{terra incognita}.

\subsection{Ginibre Orthogonal Ensemble}
\label{sec:2.GinOE}

In the development of the theory of the Gaussian and circular ensembles, the eigenvalue jpdfs (\ref{eqn:GEjpdf}) and (\ref{eqn:CEjpdf}) were contained in the pioneering papers of Dyson and Mehta \cite{dyson1962a,mehta_and_dyson1963}. In contrast, it took over 25 years from the formulation of GinOE \cite{ginibre1965} to the determination of its eigenvalue jpdf. That this problem is more complex than those solved previously can be seen from the fact that for any $N$ there is a non-zero probability of having at least one real eigenvalue --- the real line is populated despite having measure zero inside the support of the set of all eigenvalues.

The jpdf therefore breaks up into sectors labelled by the number $k$ of real eigenvalues (this number must have the same parity as the size of the matrix $N$ because the complex eigenvalues come in complex conjugate pairs) and was first computed in \cite{lehmann_and_sommers1991} and then later in \cite{edelman1997} and \cite{shukla2001}. The explicit form reads 

\begin{equation}
{\tiny
\label{eqn:GinOEjpdf}
P_{N,k}(\mathbf{\Lambda},\mathbf{W})=\frac{1}{C_{N,k}}\prod_{i=1}^{k}e^{-\lambda_i^2/2}\prod_{j=1}^{(N-k)/2} e^{-(w_j^2+\bar{w}_j^2)/2}\hspace{3pt}\mathrm{erfc}(|\mathrm{Im}(w_j)|\sqrt{2})\hspace{6pt} \Delta (\mathbf{\Lambda},\mathbf{W})}
\end{equation}
where $\mathbf{\Lambda}=\{ \lambda \}_{1,...,k}$ are the real eigenvalues and $\mathbf{W}=\{ w_i,\bar{w}_i \}_{i=1,...,(N-k)/2}$ are the complex eigenvalues, which come in conjugate pairs. $\Delta$ is the the product of absolute differences, \textit{i.e.} $\Delta(\{x_i\}):=\prod_{i<j} |x_i-x_j|$.

In any one sector $k$ the correlation function for $m$ (where $m\leq k$) real eigenvalues and $(n-m)/2$ complex eigenvalues in the upper half plane are specified by an integral analogous to (\ref{eqn:integcorrelns}), and thus recursions (one an integral over a real eigenvalue and the other an integral over a complex eigenvalue) analogous to (\ref{eqn:rho_recurrence}) hold. However, confining the correlations to a particular sector is not in keeping with the realities of the problem, in which the number of real eigenvalues is not known \textit{a priori}. Consequently, the correct way to specify the correlations is to first introduce the summed-up generating function
\begin{equation}
\label{eqn:summedup}
Z_N[u,v]=\sum_{k=0}^N{}^*Z_{k,(N-k)/2}[u,v]
\end{equation}
where $*$ indicates that the sum is restricted to values of $k$ with the same parity as $N$ and 
\begin{eqnarray}
\label{eqn:ZkN-k}
\nonumber Z_{k,(N-k)/2}[u,v]&=&\int_{-\infty}^{\infty}d\lambda_1\cdot\cdot\cdot \int_{\infty}^{\infty}d\lambda_k \prod_{l=1}^ku(\lambda_l) \\
&&\times\int_{\mathbb{R}^2_+}d\vec{w}_1\cdot\cdot\cdot\int_{\mathbb{R}^2_+}d\vec{w}_{(N-k)/2}\prod_{l=1}^{(N-k)/2}v(\vec{w}_l)\hspace{6pt} P_{N,k}(\mathbf{\Lambda},\mathbf{W})
\end{eqnarray}
with $\vec{w}_l=(x_l,y_l)$, $w_l:=x_l+iy_l$.

From (\ref{eqn:summedup}) we compute the correlation function for $m$ real eigenvalues, and $(n-m)/2$ complex eigenvalues in the upper half plane according to the functional differentiation formula
\begin{equation}
\label{eqn:GinOEfnaldiff}
\rho_{(n)}(\mathbf{x},\mathbf{w})=\frac{1}{Z_N[u,v]}\frac{\delta^n}{\delta u(x_1)\cdot\cdot\cdot \delta u(x_m)\delta v(w_1)\cdot\cdot\cdot \delta v(w_{\frac{n-m}{2}})}Z_N[u,v]{\Big |}_{u=v=1}
\end{equation}
Expanding out the functional differentiation does not lead to an integral formula for $\rho_{(n)}$ analogous to (\ref{eqn:integcorrelns}), because the RHS of such a formula would now consist of the sum of many different integrals. Thus, for the calculation of the correlations, the strategy of using (\ref{eqn:rho_recurrence}) and (\ref{eqn:rho_initial}) or their analogues, is not available for GinOE.

This leaves then just the one strategy, namely to give a Pfaffian form of the summed-up generating function (\ref{eqn:summedup}), and to deduce from this (using (\ref{eqn:1+ab}) or similar) a Pfaffian form for $\rho_{(n)}$. It was some time after the discovery of (\ref{eqn:GinOEjpdf}) before this program could be carried through. The first advance was the deduction of a Pfaffian formula for $Z_N[u,v]$ \cite{sinclair2006}
\begin{equation}
Z_N[u,v]=\frac{2^{-N(N+1)/4}}{\prod_{l=1}^{N}\Gamma(l/2)}\hspace{3pt}\mathrm{Pf}[\alpha_{j,k}[u]+\beta_{j,k}[v]]_{i,j=1,...,N}
\end{equation}
where
\begin{eqnarray}
\alpha_{j,k}[u] & = & \int_{-\infty}^{\infty}dx\hspace{3pt}u(x)\hspace{3pt}\int_{-\infty}^{\infty}dy\hspace{3pt}u(y)\hspace{3pt}e^{-(x^2+y^2)/2}p_{j-1}(x)p_{k-1}(y)\hspace{3pt}\mathrm{sgn}(y-x)\\
\nonumber \\
\beta_{j,k}[v] & = & 2i\int_{\mathbb{R}_+^2}dw\hspace{3pt}v(w)e^{y^2-x^2}\mathrm{erfc}(\sqrt{2}|\mathrm{Im}(w)|)\Bigl(p_{j-1}(w)p_{k-1}(\bar{w})-p_{k-1}(w)p_{j-1}(\bar{w})\Bigr)
\end{eqnarray}
for arbitrary monic polynomials $p_l(x)$ of degree $l$.

To proceed further, the polynomials $p_l(x)$ must be chosen to be skew-orthogonal with respect to the skew inner product
\begin{equation}
(p_{j-1},p_{k-1}):=(\alpha_j + \beta_k){\Big |}_{u=v=1}=:G_{j,k}
\end{equation}
satisfying
\begin{equation}
\label{eqn:skew-orthog_conditions}
G_{2j,2k}=G_{2j-1,2k-1}=0\hspace{6pt} ,\hspace{6pt} G_{2j-1,2k}=-G_{2k,2j-1}=r_{j-1}\delta_{j,k}
\end{equation}
The requisite polynomials are identified in \cite{forrester and nagao2007}
\begin{eqnarray}
\nonumber &&p_{2j}(x)=x^{2j}\hspace{6pt},\hspace{6pt} p_{2j+1}=x^{2j+1}-2jx^{2j-1}\\
\nonumber\\
\label{eqn:GinOE polys} &&r_{j-1}=2\sqrt{2\pi}\Gamma (2j-1)
\end{eqnarray}

With knowledge of these skew-orthogonal polynomials, it is a relatively routine task to then compute correlations using (\ref{eqn:GinOEfnaldiff}). In the case of $N$ even, through the use of (\ref{eqn:1+ab}), this leads to the computation of $\rho_{(n)}$, $n$ arbitrary, for the real-real and complex-complex correlations in \cite{forrester and nagao2007}, and then for mixed real and complex correlations in \cite{b&s2007}.

For $N$ odd it is not known how to make use of (\ref{eqn:1+ab}) (recall the remark below that formula). Nonetheless, provided $n=1,2$ the functional differentiations required by (\ref{eqn:GinOEfnaldiff}) are tractable via a different calculation. Indeed, the authors have carried through the required computations. These low order calculations, together with the structure of the general $n$-point correlations for $N$ even as known from \cite{b&s2007} immediately suggest the $n$-point correlations in the case of $N$ odd. However, what is not immediate, given the inapplicability of the recurrence (\ref{eqn:rho_recurrence}) (or an analogue) and the lack of knowledge on how to implement functional differentiation systematically for $N$ odd, is a way to verify these formulae.

This suggests revisiting the problem of computing the correlations for (\ref{eqn:generalOEjpdf}) in the case of $N$ odd. Can we devise a strategy that reclaims Theorem \ref{thm:general_odd} and will furthermore be applicable to the $N$ odd case of GinOE?

In fact this broader question is now of more importance than the calculation of the correlations for the $N$ odd case of GinOE. Very recently Sommers and Wieczorek \cite{sommers_and_w2008} studied the correlation for GinOE through functional differentiations combined with the use of Gassmannian integrals. By the use of a certain additional `artificial Grassmannian', they have solved the $N$ odd case. Our idea is quite different; we will show that the $N$ odd correlations can be obtained from the $N$ even case by a limiting procedure. Moreover, our approach is generally applicable, when given a certain determinant or Pfaffian form for the correlations pertaining to the $N$ even case.

\section{The taming of the odd}
\label{sec:odd_method}

To go from $N$ even to $N$ odd in (\ref{eqn:generalOEjpdf}) we propose taking one of the eigenvalues off to infinity. This will be a useful strategy if the jpdf exhibits the factorisation
\begin{equation}
\label{eqn:jpdfOEfactorisation}
\fullsub{P_{N,1}(x_1,...,x_N)}{\sim}{|x_1|\rightarrow\infty}{f_N(x_1)\hspace{3pt}P_{N-1,1}(x_2,...,x_N)}
\end{equation}
Indeed (\ref{eqn:generalOEjpdf}) satisfies (\ref{eqn:jpdfOEfactorisation}) with
\begin{equation}
\label{fNxOE}
f_N(x)=\frac{C_{N-1}}{C_N}x^{N-1}e^{-V(x)}
\end{equation}
It then follows from (\ref{eqn:integcorrelns}) that
\begin{equation}
\label{eqn:correlnOEfactorisation}
\fullsub{\rho_{(m)}^N(r_1,...,r_m)}{\sim}{|r_1|\rightarrow\infty}{Nf_N(r_m)\hspace{3pt}\rho_{(m-1)}^{N-1}(r_1,...,r_{m-1})}
\end{equation}
where the superscripts on the $\rho_{(k)}$ indicate the total number of eigenvalues.

With the total number fixed on the LHS, the total is reduced by one, and thus odd, on the RHS. We can further use (\ref{eqn:jpdfOEfactorisation}), (\ref{fNxOE}) and (\ref{eqn:integcorrelns}) to show
\begin{equation}
\fullsub{\rho_{(1)}^N(r)}{\sim}{|r_1|\rightarrow\infty}{Nf_N(r)}
\end{equation}
and thus rewrite (\ref{eqn:correlnOEfactorisation}) to read
\begin{equation}
\label{eqn:finalOEfactorisation}
\fullsub{\rho_{(m)}^N(r_1,...,r_m)}{\sim}{|r_1|\rightarrow\infty}{\rho_{(1)}^N(r_m)\rho_{(m-1)}^{N-1}(r_1,...,r_{m-1})}
\end{equation}

Either way, we have that the $(m-1)$-point correlation for $N$ odd is a limit of the $m$-point correlation for $N$ even. Our task now is to show how (\ref{eqn:correlnOEfactorisation}) can be used to deduce Theorem \ref{thm:general_odd} from Theorem \ref{thm:general_even}; and for that we begin with humble row and column reduction.

\subsection{Humble row and column reduction}

As discussed above, the $m$th correlations of the $\beta=1$ ensembles consist of quaternion determinants of $m\times m$ matrices with quaternion elements.

Using Theorem \ref{thm:general_even} we isolate the $m$th eigenvalue and write out the quaternion determinant explicitly, in Pfaffian form (according to (\ref{eqn:qdet=pf})).

\begin{eqnarray}
\label{eqn:Pf1}
\rho_{(m)}(x_1,...,x_m)=\mathrm{Pf}\left[\begin{array}{cc}
\left[\begin{array}{cc}
-\tilde{I}(x_i,x_j) & S(x_i,x_j)\\
-S(x_j,x_i) & D(x_i,x_j)\\
\end{array}\right] & \left[\begin{array}{cc}
-\tilde{I}(x_i,x_m) & S(x_i,x_m)\\
-S(x_m,x_i) & D(x_i,x_m)\\
\end{array}\right]\\
&\\
\left[\begin{array}{cc}
\tilde{I}(x_m,x_j) & S(x_m,x_j)\\
-S(x_j,x_m) & D(x_m,x_j)\\
\end{array}\right] & \left[\begin{array}{cc}
0 & S(x_m,x_m)\\
-S(x_m,x_m) & 0\\
\end{array}\right]\\
\end{array}\right]_{i,j=1,...,m-1}
\end{eqnarray}
for some fixed $m$. This matrix consists of 4 submatrices of sizes:
\begin{itemize}
\item{Top left: $2(m-1)\times 2(m-1)$.}
\item{Top right: $2(m-1)\times 2$.}
\item{Bottom left: $2 \times 2(m-1)$.}
\item{Bottom right: $2 \times 2$.}
\end{itemize}

\vspace{10pt}Now since $(\mathrm{Pf}[X])^2=\mathrm{Det}[X]$ for any anti-symmetric matrix $X$, (\ref{eqn:Pf1}) becomes:

\begin{equation*}
S(x_m,x_m)\mathrm{Pf}\left[\begin{array}{cc}
\left[\begin{array}{cc}
\vspace{3pt}-\tilde{I}^*(x_i,x_j) & S^*(x_i,x_j)\\
-S^*(x_j,x_i) & D^*(x_i,x_j)\\
\end{array}\right] & \left[\begin{array}{cc}
\vspace{3pt}-\tilde{I}(x_i,x_m) & 0\\
-S(x_m,x_i) & 0\\
\end{array}\right]\\
&\\
\left[\begin{array}{cc}
\vspace{3pt}-\tilde{I}(x_m,x_j) & S(x_m,x_j)\\
0 & 0\\
\end{array}\right] & \left[\begin{array}{cc}
\vspace{3pt}0 & 1\\
-1 & 0\\
\end{array}\right]\\
\end{array}\right]_{i,j=1,...,m-1}
\end{equation*}

\vspace{6pt}\begin{equation}
\label{eqn:Pf3}
=S(x_m,x_m)\mathrm{Pf}
\left[\begin{array}{cc}
\vspace{3pt}-\tilde{I}^*(x_i,x_j) & S^*(x_i,x_j)\\
-S^*(x_j,x_i) & D^*(x_i,x_j)\\
\end{array}\right]_{i,j=1,...,m-1}
\end{equation}
where\begin{eqnarray}
\label{eqn:D*}D^*(x_i,x_j)&=&D(x_i,x_j)-\frac{D(x_i,x_m)S(x_m,x_j)}{S(x_m,x_m)}-\frac{S(x_m,x_i)D(x_m,x_j)}{S(x_m,x_m)}\\
\nonumber\\
\label{eqn:S*}S^*(x_i,x_j)&=&S(x_i,x_j)-\frac{S(x_i,x_m)S(x_m,x_j)}{S(x_m,x_m)}-\frac{D(x_m,x_j)\tilde{I}(x_i,x_m)}{S(x_m,x_m)}\\
\nonumber\\
\label{eqn:I*}\tilde{I}^*(x_i,x_j)&=&\tilde{I}(x_i,x_j)-\frac{S(x_i,x_m)\tilde{I}(x_m,x_j)}{S(x_m,x_m)}-\frac{S(x_j,x_m)\tilde{I}(x_i,x_m)}{S(x_m,x_m)}
\end{eqnarray}

The equality in (\ref{eqn:Pf3}) can be seen by using the Laplace expansion method for Pfaffians discussed in Section \ref{subsec:QDet and Pf}. Note that (\ref{eqn:Pf3}) factors out $\rho_{(1)}^N$ as required by (\ref{eqn:finalOEfactorisation}).

\subsection{To infinity and beyond}

To compute $\rho_{(m-1)}^{N-1}$ according to (\ref{eqn:finalOEfactorisation}), we must cancel $S(x_m,x_m)$ and take $x_m\rightarrow\infty$ in the entries of the remaining Pfaffian.

To reclaim Theorem \ref{thm:general_odd} we must then have:

\begin{eqnarray}
\label{eqn:D*=Dodd}&&\fullsub{D^*(x_i,x_j){\Big |}_{N\rightarrow N-1}}{\sim}{x_m\rightarrow\infty}{D^{odd}(x_i,x_j)}\\
&&\nonumber\\
\label{eqn:S*=Sodd}&&\fullsub{S^*(x_i,x_j){\Big |}_{N\rightarrow N-1}}{\sim}{x_m\rightarrow\infty}{S^{odd}(x_i,x_j)}\\
&&\nonumber\\
\label{eqn:I*=Iodd}&&\fullsub{\tilde{I}^*(x_i,x_j){\Big |}_{N\rightarrow N-1}}{\sim}{x_m\rightarrow\infty}{\tilde{I}^{odd}(x_i,x_j)}
\end{eqnarray}
Now using (\ref{eqn:Seven}), (\ref{eqn:Deven}) and (\ref{eqn:Ieven}) we see that as $x_m\rightarrow \infty$

\begin{eqnarray}
\label{eqn:Dxm}D(x_i,x_m)&\rightarrow &\frac{e^{-V(x_i)}e^{-V(x_m)}}{r_{N/2-1}}R_{N-2}(x_i)\hspace{3pt}R_{N-1}(x_m)\\
\nonumber\\
\label{eqn:Sxm1}S(x_i,x_m)&\rightarrow &\frac{e^{-V(x_m)}}{r_{N/2-1}}\Phi_{N-2}(x_i)\hspace{3pt}R_{N-1}(x_m)\\
\nonumber\\
\nonumber S(x_m,x_i)&\rightarrow &\sum_{k=0}^{N/2-1}\frac{e^{-V(x_i)}}{r_k}\Bigl[R_{2k+1}(x_i)\hspace{3pt} \frac{1}{2}\int_{\infty}^{\infty}e^{-V(y)}R_{2k}(y)dy\\
\label{eqn:Sxm2}&&- R_{2k}(x_i)\hspace{3pt} \frac{1}{2}\int_{\infty}^{\infty}e^{-V(y)}R_{2k+1}(y)dy\Bigr]\\
\nonumber\\
\label{eqn:Sxmxm}S(x_m,x_m)&\rightarrow &\frac{e^{-V(x_m)}}{r_{N/2-1}}R_{N-1}(x_m)\hspace{3pt}\frac{1}{2}\int_{\infty}^{\infty}e^{-V(y)}R_{N-2}(y)dy\\
\nonumber\\
\nonumber \tilde{I}(x_i,x_m)&\rightarrow &\sum_{k=0}^{N/2-1}\frac{1}{r_k}\Bigl[\Phi_{2k+1}(x_i)\hspace{3pt} \frac{1}{2}\int_{\infty}^{\infty}e^{-V(y)}R_{2k}(y)dy\\
\label{eqn:Ixm}&&- \Phi_{2k}(x_i)\hspace{3pt} \frac{1}{2}\int_{\infty}^{\infty}e^{-V(y)}R_{2k+1}(y)dy\Bigr]+\frac{1}{2}
\end{eqnarray}

From these we find the equalities (\ref{eqn:D*=Dodd}), (\ref{eqn:S*=Sodd}) and (\ref{eqn:I*=Iodd}), and so Theorem \ref{thm:general_odd} is reclaimed.

\section{GinOE into the fold}

Here we show that GinOE is amenable to the same treatment; we claim that an analogy of the asymptotic factorisation formula (\ref{eqn:finalOEfactorisation}) again holds true. To understand this we return to the jpdf (\ref{eqn:GinOEjpdf}) and impose the ordering constraint
\begin{equation}
\label{eqn:orderedEvals}
\lambda_1 < \lambda_2 < \cdot\cdot\cdot < \lambda_k
\end{equation}

One significance of this is that the normalisation $C_{N,k}$, which, prior to this ordering, is dependent on $k$ through the factor $k!$ , is then $C_{N,k}=C_{N,N-k}$, a function of $N$ and $N-k$ only \cite{lehmann_and_sommers1991,edelman1997}. The formula (\ref{eqn:ZkN-k}) must correspondingly be modified in its domains of integration to account for these orderings. With this understood, we observe that for large $x$
\begin{equation}
\label{eqn:ZdiffInfinity}
\fullsub{\frac{\delta}{\delta u(x)}Z_{k,(N-k)/2}[u,v]}{\sim}{x\rightarrow\infty}{\frac{C_{N-1,N-k}}{C_{N,N-k}}\hspace{3pt}e^{-x^2/2}x^{N-1}\hspace{3pt}Z_{k-1,(N-k)/2}[u,v]}
\end{equation}

The reason for this is that in this limit the leading contribution comes from the functional derivative acting on the $\lambda_k$ variable as ordered in (\ref{eqn:orderedEvals}). But according to (\ref{eqn:GinOEjpdf}), and the remark made above relating to $C_{N,k}$, the jpdf contributing to each term in (\ref{eqn:summedup}) keeps its functional form, but with $N\mapsto N-1$ and $k \mapsto k-1$ and an overall factor of $\frac{C_{N-1,N-k}}{C_{N,N-k}}\hspace{3pt}e^{-x^2/2}x^{N-1}$.

This latter factor is the leading $x\rightarrow \infty$ form of $\rho_{(1)}(x)$, as can be seen from (\ref{eqn:GinOEfnaldiff}) with $n=m=1$ and (\ref{eqn:ZdiffInfinity}) itself. Use of (\ref{eqn:ZdiffInfinity}) in (\ref{eqn:GinOEfnaldiff}) for general $n$ gives as the analogue of (\ref{eqn:finalOEfactorisation}) for GinOE
\begin{equation}
\label{eqn:finalGinOEfactorisation}
\fullsub{\rho_{(n_1,n_2)}^N(\mathbf{x},\mathbf{w})}{\sim}{x_1\rightarrow\infty}{\rho_{(1,0)}^N(x_1)\rho_{(n_1-1,n_2)}^{N-1}(\mathbf{x},\mathbf{w})}
\end{equation}
where $\rho^N_{(n_1,n_2)}$ is the correlation function for $n_1$ real and $n_2$ complex eigenvalues from the ensemble of $N\times N$ matrices.

To proceed, we first need the appropriate even solution; for this we use that of \cite{b&s2008}.

\begin{theorem}

Let $\{p_i\}_{i=1,...}$ be the skew-orthogonal polynomials of (\ref{eqn:GinOE polys}) then 
\begin{equation}
\label{eqn:GinOE_even}
\rho^N_{(n_1,n_2)}(\mathbf{x},\mathbf{w})=\mathrm{Pf}\left[\begin{array}{cc}
K_N(x_i,x_j) & K_N(x_i,w_l)\\
K_N(w_k,x_j) & K_N(w_k,w_l)\\
\end{array}\right]\hspace{10pt} x_i\in \mathbb{R}\hspace{3pt}, \hspace{3pt} w_i \in \mathbb{R}_2^+
\end{equation}
\begin{equation}
\label{eqn:GinOE_kernel}
K_N(s,t)=\left[\begin{array}{cc}
D(s,t) & S(s,t)\\
-S(t,s) & I(s,t)\\
\end{array}\right]
\end{equation}
for$$\mathbf{x}=\{ x_1,...,x_{n_1} \} \hspace{6pt},\hspace{6pt} \mathbf{w}=\{ w_1,...,w_{n_2} \}$$
where

\begin{eqnarray}
D(\mu,\eta)&=&2\sum_{k=0}^{\frac{N}{2}-1}\frac{1}{r_k}\Bigl[q_{2k}(\mu)q_{2k+1}(\eta)-q_{2k+1}(\mu)q_{2k}(\eta)\Bigr]\\
S(\mu,\eta)&=&2\sum_{k=0}^{\frac{N}{2}-1}\frac{1}{r_k}\Bigl[q_{2k}(\mu)\tau_{2k+1}(\eta)-q_{2k+1}(\mu)\tau_{2k}(\eta)\Bigr]\\
I(\mu,\eta)&=&2\sum_{k=0}^{\frac{N}{2}-1}\frac{1}{r_k}\Bigl[\tau_{2k}(\mu)\tau_{2k+1}(\eta)-\tau_{2k+1}(\mu)\tau_{2k}(\eta)\Bigr]+\epsilon(\mu,\eta)
\end{eqnarray}
and

\begin{eqnarray}
q_i(z) &=& e^{z^2/2}\hspace{2pt}\sqrt{\mathrm{erfc}(\sqrt{2}|\mathrm{Im}(z)|)}\hspace{2pt}p_i
(z)\\
\nonumber\\
\tau_i(z) &=& 
\left\{ 
\begin{array}{ll}
ie^{\bar{z}^2/2}\hspace{2pt}\sqrt{\mathrm{erfc}(\sqrt{2}|\mathrm{Im}(z)|)}\hspace{2pt}p_i(\bar{z})  & z\in \mathbb{R}_2^+\\
-\frac{1}{2}\Phi_i(z)  & z\in \mathbb{R}\\
\end{array}
\right.\\
\nonumber\\
\epsilon(z_1,z_2) &=& 
\left\{ 
\begin{array}{ll}
\frac{1}{2}\mathrm{sgn}(z_1-z_2)  & z_1,z_2\in \mathbb{R}\\
0  & \mathrm{otherwise}\\
\end{array}
\right.\\
\nonumber\\
\Phi_{j}(x)&=&\int_{-\infty}^{\infty}\mathrm{sgn}(x-z)\hspace{3pt}p_j(z)e^{-z^2/2}\hspace{3pt}dz
\end{eqnarray}

\end{theorem}

\vspace{26pt}The matrix in (\ref{eqn:GinOE_even}) consists of four blocks representing the four possible combinations of eigenvalues: real--real, real--complex, complex--real and complex--complex. For ease of manipulation we shift the $m$th real eigenvalue to the right-most column and the bottom row. Since we are shifting two rows and columns an even number of times we do not change the Pfaffian. We now have:

\begin{equation}
\mathrm{Pf}\left[\begin{array}{ccc}
\vspace{6pt}K_N(x_i,x_j) & K_N(x_i,w_l) & K_N(x_i,x_m)\\
\vspace{6pt}K_N(w_k,x_j) & K_N(w_k,w_l) & K_N(w_k,x_m)\\
\vspace{6pt}K_N(x_m,x_j) & K_N(x_m,w_l) & K_N(x_m,x_m)\\
\end{array}\right]
\end{equation}

The sizes of the submatrices are:

\begin{itemize}
\item{top left: $2(m-1)\times 2(m-1)$; top centre: $2(m-1)\times 2(n-m)$; top right: $2(m-1)\times 2$.}
\item{centre left: $2(n-m)\times 2(m-1)$; centre: $2(n-m)\times 2(n-m)$; centre right: $2(n-m)\times 2$.}
\item{bottom left: $2\times 2(m-1)$; bottom centre: $2\times 2(n-m)$; bottom right: $2\times 2$.}
\end{itemize}

It can be seen that this matrix is now equivalent to that in Equation (\ref{eqn:Pf1}). By applying the same process as in Section \ref{sec:odd_method} above, we arrive at the odd case. The details of the calculation involve index-shuffling with a liberal coating of tedium and so they are omitted. The results, however, are:

\begin{theorem}
\label{thm:GinOE_odd}

For $N$ odd, (\ref{eqn:GinOE_even}) and (\ref{eqn:GinOE_kernel}) hold with the following modifcations
\begin{eqnarray}
D(\mu,\eta)&=&2\sum_{k=0}^{\frac{(N-1)}{2}-1}\frac{1}{r_k}\Bigl[\hat{q}_{2k}(\mu)\hat{q}_{2k+1}(\eta)-\hat{q}_{2k+1}(\mu)\hat{q}_{2k}(\eta)\Bigr]\\
\nonumber\\
S(\mu,\eta)&=&2\sum_{k=0}^{\frac{(N-1)}{2}-1}\frac{1}{r_k}\Bigl[\hat{q}_{2k}(\mu)\hat{\tau}_{2k+1}(\eta)-\hat{q}_{2k+1}(\mu)\hat{\tau}_{2k}(\eta)\Bigr]+\kappa(\mu,\eta)\\
\nonumber\\
I(\mu,\eta)&=&2\sum_{k=0}^{\frac{(N-1)}{2}-1}\frac{1}{r_k}\Bigl[\hat{\tau}_{2k}(\mu)\hat{\tau}_{2k+1}(\eta)-\hat{\tau}_{2k+1}(\mu)\hat{\tau}_{2k}(\eta)\Bigr]+\epsilon(\mu,\eta)+\theta(\mu,\eta)
\end{eqnarray}
where
\begin{eqnarray}
\hat{q}_i(z) &=& e^{z^2/2}\hspace{2pt}\sqrt{\mathrm{erfc}(\sqrt{2}|\mathrm{Im}(z)|)}\hspace{2pt}\hat{p}_i(z)\\
\nonumber\\
\hat{\tau}_i(z) &=& 
\left\{ 
\begin{array}{ll}
ie^{\bar{z}^2/2}\hspace{2pt}\sqrt{\mathrm{erfc}(\sqrt{2}|\mathrm{Im}(z)|)}\hspace{2pt}\hat{p}_i(\bar{z})  & z\in \mathbb{R}_2^+\\
\\
-\frac{1}{2}\Bigl(\Phi_i(z)-\frac{\nu_{i+1}}{\nu_N}\Phi_{N-1}(z)\Bigr)  & z\in \mathbb{R}\\
\end{array}
\right.\\
\nonumber\\
\kappa(\mu,\eta) &=& 
\left\{ 
\begin{array}{ll}
\vspace{3pt}\frac{1}{2\nu_N}\hspace{2pt}e^{\mu^2/2}\sqrt{\mathrm{erfc}(\sqrt{2}|\mathrm{Im}(\mu)|)}\hspace{2pt}p_{N-1}(\mu)  & \eta\in \mathbb{R}\\
0  & \eta\in \mathbb{R}_2^+\\
\end{array}
\right.\\
\nonumber\\
\theta(\mu,\eta) &=& 
\left\{ 
\begin{array}{ll}
\vspace{3pt}\frac{1}{4\nu_N}\Bigl(\Phi_{N-1}(\eta)-\Phi_{N-1}(\mu)\Bigr) & \mu,\eta\in \mathbb{R}\\
\vspace{3pt}-\frac{1}{2\nu_N}\hspace{2pt}\tau_{N-1}(\eta) & \mu\in\mathbb{R}\hspace{3pt},\hspace{2pt} \eta\in \mathbb{R}_2^+\\
\vspace{3pt}\frac{1}{2\nu_N}\hspace{2pt}\tau_{N-1}(\mu) & \mu\in\mathbb{R}_2^+\hspace{3pt},\hspace{2pt} \eta\in \mathbb{R}\\
0  & \mu,\eta\in \mathbb{R}_2^+\\
\end{array}
\right.\\
\nonumber\\
\hat{p}_i(z)&=&p_i(z)-\frac{\nu_{i+1}}{\nu_N}p_{N-1}(z)\hspace{16pt}i=1,...N-2\\
\nonumber\\
\nu_l&=&\frac{1}{2}\int_{-\infty}^{\infty}u(\lambda)e^{-\lambda^2/2}p_{l-1}(\lambda)\hspace{3pt}d\lambda
\end{eqnarray}
\end{theorem}

\vspace{26pt} This is the exact form of the $N$ odd correlation functions for GinOE. However, there are still two further points to appreciate. One is that there are inter-relations between the $D,S,I$ within the various blocks, \textit{i.e.} for the real-real correlations, complex-complex correlations, etc. For the real-real case the relationships are presented in \cite{forrester and nagao2007}, they amount to differentiation and integration of $S_{r,r}(x,y)$ to obtain $D_{r,r}(x,y)$ and $I_{r,r}(x,y)$ respectively. In the complex-complex case, the operation is complex conjugation of one of the variables and multiplication by $i$. The mixed cases are a combination of these operations. The inter-relationships are the same for both $N$ even and $N$ odd.

The other point is that there exist summed-up forms of, say, the $S$ in each block, and thus the $D$ and $I$ by the inter-relations, which are applicable for both $N$ even and $N$ odd. In the case of $S_{r,r}(x,x)$, this has been known since the work of Edelman \textit{et al}. \cite{eks1994}, and has been shown to be true in general in \cite{sommers_and_w2008}.

We collect together the formulae bearing on these points in the Appendix.

\section{Concluding remarks}

The problem of computing correlation functions for GinOE is complicated by the eigenvalue jpdf breaking up into sectors according to the number of real eigenvalues. It is further obscured by the need to treat the $N$ even and $N$ odd cases separately. This latter complication is shared by the general $\beta=1$ eigenvalue jpdf (\ref{eqn:generalOEjpdf}). It has motivated us to revisit the problem of computing correlations in the $N$ odd case for (\ref{eqn:generalOEjpdf}), with our aim being to devise a method applicable to the GinOE for $N$ odd. We explain, in Sections \ref{sec:guess} and \ref{sec:2.GinOE}, why the established method for computing correlations for (\ref{eqn:generalOEjpdf}) in the case $N$ odd cannot be applied to GinOE.

The approach we take is to compute the $N$ odd correlations as limiting cases of the $N$ even correlations according to (\ref{eqn:finalOEfactorisation}), in relation to (\ref{eqn:generalOEjpdf}), and (\ref{eqn:finalGinOEfactorisation}) in relation to GinOE. Pfaffian forms of the appropriate size are obtained from these by row and column reductions of Pfaffian form implied by the LHSs.

We remark that (\ref{eqn:finalOEfactorisation}) and (\ref{eqn:finalGinOEfactorisation}) work equally well for relating the correlations for $N$ odd to those for $N$ even, which can be used as a consistency check on our own workings. Further, there are other problems to which our method applies. The most immediate is the partially symmetric GinOE \cite{lehmann_and_sommers1991,forrester_and_nagao2008}, which will be the topic of a future publication.

\section*{Acknowledgements}
The work of PJF was supported by the Australian Research Council,
and AM was supported by an Australian Postgraduate Award.

\newpage

\appendix\section*{Appendix}
\setcounter{equation}{0}
\renewcommand{\theequation}{A.\arabic{equation}}

The explicit forms of the terms in the correlation kernel (\ref{eqn:GinOE_kernel}) for $N$ odd are contained below. The subscripts $r$ and $c$ identify real--real, real--complex, complex--real and complex-complex factors. The convention used in this Appendix is $x,y\in\mathbb{R}$ and $w,z\in\mathbb{R}_2^+$.

As mentioned after Theorem \ref{thm:GinOE_odd} there are inter-relationships between the various kernel elements as seen in \cite{forrester and nagao2007,b&s2007,sommers_and_w2008}. Note the interesting fact that to obtain $I$ from $S$ you operate on the first variable, and to obtain $D$ from $S$ you operate on the second variable.

\begin{eqnarray}
I_{r,r}(x,y)&=&-\int_{x}^{y}S_{r,r}(z,y)dz+\frac{1}{2}\mathrm{sgn}(x-y)\\
\nonumber&&\\
D_{r,r}(x,y)&=&-\frac{\partial}{\partial y}S_{r,r}(x,y)\\
\nonumber&&\\
I_{r,c}(x,w)&=&-I_{c,r}(w,x)= -\int_{x}^{y}S_{r,c}(z,w)dz\\
\nonumber&&\\
D_{r,c}(x,w)&=&-D_{c,r}(w,x)=-iS_{r,c}(x,\bar{w})\\
\nonumber&&\\
I_{c,r}(w,x)&=&-I_{r,c}(x,w)=iS_{c,r}(\bar{w},x)\\
\nonumber&&\\
D_{c,r}(w,x)&=&-D_{r,c}(x,w)=-\frac{\partial}{\partial x}S_{c,r}(w,x)\\
\nonumber&&\\
I_{c,c}(w,x)&=&iS_{c,c}(\bar{w},z)\\
\nonumber&&\\
D_{c,c}(w,z)&=&-iS_{c,c}(w,\bar{z})
\end{eqnarray}

Below are listed the kernel elements $S$ for real--real, real--complex, complex--real and complex--complex correlations. Using the above relationships, the other elements can be deduced.

\begin{eqnarray}
\nonumber S_{r,r}(x,y)&=&e^{-x^2/2}\sum_{k=0}^{\frac{N-1}{2}-1}\frac{1}{r_k}\Bigl[p_{2k+1}(x)\Phi_{2k}(y)-p_{2k}(x)\Phi_{2k+1}(y)\\
\nonumber &&-\frac{\nu_{2k+1}}{\nu_N}\Bigl(p_{2k+1}(x)\Phi_{N-1}(y)-p_{N-1}(x)\Phi_{2k+1}(y)\Bigr)\\
\nonumber &&+\frac{\nu_{2k+2}}{\nu_N}\Bigl(p_{2k}(x)\Phi_{N-1}(y)-p_{N-1}(x)\Phi_{2k}(y)\Bigr)\Bigr]\\
&&+\frac{e^{-x^2/2}}{2\nu_N}p_{N-1}(x)\\
\nonumber &&\\
\nonumber S_{r,c}(x,w)&=&2ie^{-x^2/2}e^{-\bar{w}^2/2}\sqrt{\mathrm{erfc}(\sqrt{2}|\mathrm{Im}(w)|)}\\
\nonumber &&\times\sum_{k=0}^{\frac{N-1}{2}-1}\frac{1}{r_k}\Bigl[p_{2k}(x)p_{2k+1}(\bar{w})-p_{2k+1}(x)p_{2k}(\bar{w})\\
\nonumber &&-\frac{\nu_{2k+1}}{\nu_N}\Bigl(p_{N-1}(x)p_{2k+1}(\bar{w})-p_{2k+1}(x)p_{N-1}(\bar{w})\Bigr)\\
&&+\frac{\nu_{2k+2}}{\nu_N}\Bigl(p_{N-1}(x)p_{2k}(\bar{w})-p_{2k}(x)p_{N-1}(\bar{w})\Bigr)\Bigr]\\
\nonumber &&\\
\nonumber S_{c,r}(w,x)&=&e^{-w^2/2}\sqrt{\mathrm{erfc}(\sqrt{2}|\mathrm{Im}(w)|)}\\
\nonumber &&\times\sum_{k=0}^{\frac{N-1}{2}-1}\frac{1}{r_k}\Bigl[p_{2k+1}(w)\Phi_{2k}(x)-p_{2k}(w)\Phi_{2k+1}(x)\\
\nonumber &&-\frac{\nu_{2k+1}}{\nu_N}\Bigl(p_{2k+1}(w)\Phi_{N-1}(x)-p_{N-1}(w)\Phi_{2k+1}(x)\Bigr)\\
\nonumber &&+\frac{\nu_{2k+2}}{\nu_N}\Bigl(p_{2k}(w)\Phi_{N-1}(x)-p_{N-1}(w)\Phi_{2k}(x)\Bigr)\Bigr]\\
&&+\frac{e^{-w^2/2}}{2\nu_N}\sqrt{\mathrm{erfc}(\sqrt{2}|\mathrm{Im}(w)|)}\hspace{3pt}p_{N-1}(w)\\
\nonumber &&\\
\nonumber S_{c,c}(w,z)&=&2ie^{-w^2/2}e^{-\bar{z}^2/2}\sqrt{\mathrm{erfc}(\sqrt{2}|\mathrm{Im}(w)|)}\sqrt{\mathrm{erfc}(\sqrt{2}|\mathrm{Im}(z)|)}\\
\nonumber &&\times\sum_{k=0}^{\frac{N-1}{2}-1}\frac{1}{r_k}\Bigl[p_{2k}(w)p_{2k+1}(\bar{z})-p_{2k+1}(w)p_{2k}(\bar{z})\\
\nonumber && -\frac{\nu_{2k+1}}{\nu_N}\Bigl(p_{N-1}(w)p_{2k+1}(\bar{z})-p_{2k+1}(w)p_{N-1}(\bar{z})\Bigr)\\
&&+\frac{\nu_{2k+2}}{\nu_N}\Bigl(p_{N-1}(w)p_{2k}(\bar{z})-p_{2k}(w)p_{N-1}(\bar{z})\Bigr)\Bigr]
\end{eqnarray}

The kernel elements $S$ can also be summed-up \cite{forrester_and_nagao2008,sommers_and_w2008}; the summed up forms are listed here. Note that these formulae are insensitive to the parity of $N$ (in fact $N$ need not even be an integer), and so provide a further check of our work in this paper.

\begin{eqnarray}
S_{r,r}(x,y)&=&\frac{1}{\sqrt{2\pi}}\Bigl[e^{-(x-y)^2/2}\hspace{3pt}\frac{\Gamma(N-1,xy)}{\Gamma(N-1)}-2^{(N-3)/2}\hspace{3pt}e^{-x^2/2}\hspace{3pt}x^{N-1}\frac{\gamma(\frac{N-1}{2},y^2/2)}{\Gamma(N-1)}\Bigr]\\
\nonumber&&\\
S_{r,c}(x,w)&=&\frac{i\hspace{1pt}e^{-(x-\bar{w})^2/2}}{\sqrt{2\pi}}(\bar{w}-x)\frac{\Gamma(N-1,x\bar{w})}{\Gamma(N-1)}\sqrt{\mathrm{erfc}(\sqrt{2}|\mathrm{Im}(w)|)}\\
\nonumber &&\\
\nonumber S_{c,r}(w,x)&=&\frac{1}{\sqrt{2\pi}}\Bigl[e^{-(w-x)^2/2}\hspace{3pt}\frac{\Gamma(N-1,wx)}{\Gamma(N-1)}-2^{(N-3)/2}\hspace{3pt}e^{-w^2/2}\hspace{3pt}w^{N-1}\frac{\gamma(\frac{N-1}{2},x^2/2)}{\Gamma(N-1)}\Bigr]\\
&&\times\sqrt{\mathrm{erfc}(\sqrt{2}|\mathrm{Im}(w)|)}\\
\nonumber&&\\
S_{c,c}(w,z)&=&\frac{i\hspace{1pt}e^{-(w-\bar{z})^2/2}}{\sqrt{2\pi}}(\bar{z}-w)\frac{\Gamma(N-1,w\bar{z})}{\Gamma(N-1)}\sqrt{\mathrm{erfc}(\sqrt{2}|\mathrm{Im}(w)|)}\sqrt{\mathrm{erfc}(\sqrt{2}|\mathrm{Im}(z)|)}
\end{eqnarray}

\end{document}